# Thermodynamic Self-Assembly of Two-Dimensional π-Conjugated Metal–Porphyrin Covalent Organic Frameworks by "On-Site" Equilibrium Polymerization

Ryota Tanoue[1], Rintaro Higuchi[1], Kiryu Ikebe[1], Shinobu Uemura[1], Nobuo Kimizuka[2,3], Adam Z. Stieg[4,5], James K. Gimzewski[4,5,6], Masashi Kunitake[1,2]*

[1]Graduate School of Science and Technology, Kumamoto University, 2-39-1 Kurokami, Chuo-ku, Kumamoto 860-8555, Japan
[2]Core Research for Evolutional Science and Technology, Japan Science and Technology Agency (JST-CREST), Kawaguchi Center Building, 4-1-8 Honcho, Kawaguchi, Saitama 332-0012, Japan
[3]Department of Chemistry and Biochemistry, Graduate School of Engineering, International Research Center for Molecular Systems (IRCMS), Kyushu University, 744 Moto-oka, Nishi-ku, Fukuoka 819-0395, Japan
[4]California NanoSystems Institute, 570 Westwood Plaza, Los Angeles, CA 90095, USA
[5]WPI Center for Materials Nanoarchitectonics (MANA), National Institute for Materials Science (NIMS), 1-1 Namiki, Tsukuba, Ibaraki 305-0044, Japan
[6]Department of Chemistry and Biochemistry, University of California-Los Angeles, 607 Charles E. Young Drive East, Los Angeles, CA 90095, USA

**ABSTRACT.** Two-dimensional π-conjugated metal–porphyrin covalent organic frameworks were produced in aqueous solution on an iodine-modified Au(111) surface by "on-site" azomethine coupling of $Fe^{III}$-5,10,15,20-tetrakis(4-aminophenyl)porphyrin (FeTAPP) with terephthal dicarboxaldehyde and investigated in detail using *in-situ* scanning tunneling microscopy. Mixed covalent organic porphyrin frameworks consisting of FeTAPP and metal-free TAPP ($H_2$TAPP) were prepared through simultaneous adsorption in a mixed solution as well as partial replacement of FeTAPP by $H_2$TAPP in an as-prepared metal-porphyrin framework. In the mixed framework, the relative distribution of FeTAPP to $H_2$TAPP was not random and revealed a preference for homo-connection rather than hetero-connection. The construction of substrate-supported, π-conjugated covalent frameworks from multiple building blocks, including metal centers, will be of significant utility in the design of functional molecular nanoarchitectures.

## 1. INTRODUCTION

Highly ordered macromolecular systems with sophisticated periodic structures extended in two or three dimensions (2-D or 3-D) represent an emerging class of potentially impactful materials. Molecular and supramolecular assemblies based on relatively weak intermolecular interactions such as hydrogen bonding have been shown to be effective in the production of such extended, periodic structures[1-3]. More recently, there has been an increased focus on methods for preparing porous coordination polymers. Metal–organic frameworks (MOFs)[4-8] based on metal–ligand complexation have attracted much attention because of their synthetic flexibility, desirable properties, and successful implementation in gas-storage[9,10], separation, photonic, optoelectronic[11-13], and catalytic applications[14]. Despite substantial successes in metal-coordination chemistry over the last 20 years, the stability limitations inherent to dipolar bonding have stimulated efforts toward the design and synthesis of periodic crystalline structures which exploit the strength of covalent bonds between organic molecular building blocks, i.e., covalent organic frameworks (COFs).[15,16] To date, this burgeoning field has produced a variety of exciting 2-D and 3-D COFs. However, the majority of synthetic methods for the production of these materials generate insoluble microcrystalline powders[13,15-21] that are unsuitable for various applications of interest.[22]

Substrate-supported molecular frameworks are of particular interest in both monolayer and thin-film forms. The vast majority of effort to this end has involved construction of 2-D molecular assemblies and monolayers on well-defined surfaces under ultrahigh vacuum (UHV) conditions, with more limited success in organic and aqueous solutions. Through enabling studies of on-site covalent bond formation, thermally initiated C-C coupling[23-25], Ullmann coupling between aromatic halogen molecules[26-28], esterification between boronic acid and hydroxyl units[29,30], and azomethine coupling between primary amine and aldehyde units[31,32] have all been reported under UHV conditions. Using more simple thermal vapor treatments, a massive 2-D honeycomb structure consisting of boronic acid dehydrate has been achieved.[33,34] In addition, chain polymerization of adsorbed monomers such as thiophene and diacetylene analogs, initiated by tip-induced bias[35,36] and electrochemical pulses[37,38] have demonstrated the feasibility of applying on-site polymerization in ambient and liquid environments. To bridge the gap between bulk production of microcrystalline powders and monolayer preparations, efforts to prepare multilayers and ultrathin molecular films have resulted in the production of ultrathin MOF films by successive deposition at air–water interfaces[39-41] and stepwise layer-by-layer growth[42-45], as well as ultrathin boronic COF films incubated under solvothermal conditions.[46]

*Corresponding author: kunitake@kumamoto-u.ac.jp

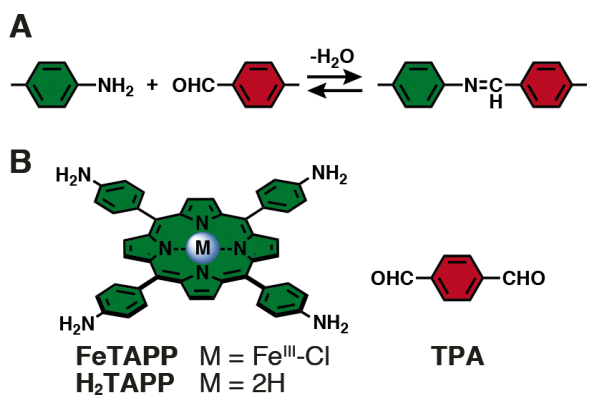

**Fig. 1.** Reaction scheme of azomethine coupling, and chemical structures of building blocks used in this work.

Recently, we reported a soft solution methodology for the preparation of extended π-conjugated polymeric nanoarchitectures in which the building blocks connected with each other via π-conjugated azomethine bonds (Figure 1A).[47,48] Self-assembly of sophisticated covalent nanostructures on well-defined surfaces in aqueous solution is achieved by reversible equilibrium polymerization in a similar manner to thermodynamically controlled, non-covalent self-assembly based on weak interactions, such as hydrogen bonds. Based on thermodynamic control of equilibrium polymerization at the solid–liquid interface, whereby aromatic building blocks spontaneously and selectively form linkages, close-packed arrays composed of one-dimensional (1-D) aromatic polymers and two-dimensional (2-D) macromolecular frameworks consisting of metal-free porphyrin and bifunctional aromatic linkage molecules have been prepared and observed *in situ* by scanning tunneling microscopy (STM).

Careful regulation of the solution conditions, primarily the pH, provides control over the moderate adsorption/partition and reaction equilibriums. Reactions in the solution phase are thus essentially prohibited and can proceed only on the substrate, where hydrophobic and chemically inert surfaces show distinct advantages over reactions in homogeneous solutions. Further, a new methodology based on these same reactions and philosophy, enabled the successive deposition of polyazomethine multilayers through chemical liquid deposition by substrate-assisted polycondensation, whereby successive depositions can be achieved by slight promotion of the reaction equilibrium.[49]

Here, we build upon these prior results to introduce a π-conjugated 2-D covalent organic framework composed of a metal–porphyrin [$Fe^{III}$-5,10,15,20-tetrakis(4-aminophenyl)porphyrin; FeTAPP] and mixtures of FeTAPP and metal-free TAPP ($H_2$TAPP) produced by surface-selective "on-site" azomethine coupling reactions with terephthaldicarboxaldehyde (TPA; Figure1B).

## 2. EXPERIMENTAL

### 2.1 Chemicals

All chemicals were purchased from commercial suppliers and used without further purification.

### 2.2 Sample preparation

Sample solutions of porphyrins and aromatic aldehydes were dissolved in aqueous solutions of 80 mM $NaClO_4$ (85%; Katayama Chemical, Osaka, Japan). The solution pH was then carefully adjusted by addition of aqueous solutions of $HClO_4$ (60%; Kanto Chemical, Tokyo, Japan) and/or NaOH. Typical concentrations of FeTAPP (Porphyrin Systems, Appen, Germany), $H_2$TAPP (TCI, Tokyo, Japan), and TPA (95%; TCI Tokyo, Japan) were 1.6–8.0, 0.50–0.65, and 30 $\mu$M, respectively. Because of the low solubility of FeTAPP in pure water, FeTAPP was dissolved in 1.0 M $HClO_4$ and added to 80 mM $NaClO_4$ solution prior to adjusting the solution pH to 3.5±0.2.

### 2.3 In-situ STM observations

Au(111) single-crystal bead electrodes for STM observations were prepared from Au wire (99.999%; Tanaka Kikinzoku, Tokyo, Japan) by the Clavilier method, according to a previous report.[50] Iodine-modified Au(111) electrodes were prepared by immersion in an aqueous solution containing 10 mM KI (99.5%; Kanto Chemical, Tokyo, Japan) for a few seconds. Details of the preparation of an iodine-modified Au(111) substrate have been given in previous articles.[51-53] STM observations were carried out using a Nanoscope E microscope (Digital Instruments, Santa Barbara, CA, USA) and all STM images were collected in constant-current mode. The iodine-modified Au(111) electrode was transferred into an STM cell filled with pH-controlled reaction solutions. Tungsten wires (Niraco, Tokyo, Japan) electrochemically-etched in 1.0 M KOH (85%; Nacalai Tesque, Kyoto, Japan) were used as STM tips. Tips were coated with clear commercial nail polish to minimize the faradaic current. In addition, 10 $\mu$L of $H_2$TAPP solution (81 $\mu$M) were added to the STM cell during operation after the self-assembly of FeTAPP, in order to observe the replacement of porphyrin molecules.

## 3. RESULTS AND DISCUSSION

### 3.1 Construction of FeTAPP–TPA mesh

Preparation of highly ordered homo-molecular mesh adlayers on I/Au(111) was confirmed by STM imaging as shown in Figure 2. The 2D covalently-bonded supra-macromolecular framework consisting of FeTAPP was observed to be constructed in an identical fashion to mesh adlayers of metal-free $H_2$TAPP. Island-like domain structures ranging in size from 20 to 50 nm$^2$, comprised of ideal 2D square mesh structures, covered roughly 70% the surface as shown in Figure 2. At this level of surface coverage, a few irregular structures such as a linearly connected porphyrin chains and arrays of chains were found. As mentioned in a previous report[47], such irregular structures were frequently observed in cases of higher surface coverage resulting from thermodynamic selectivity of higher packing density in linearly connected porphyrins. In the FeTAPP–TPA system, a porphyrin concentration approximately three times higher than in the case of $H_2$TAPP–TPA was required to achieve an ordered mesh structure with a similar surface coverage, a behavior indicative of weaker interactions between FeTAPP and the I/Au(111) substrate.

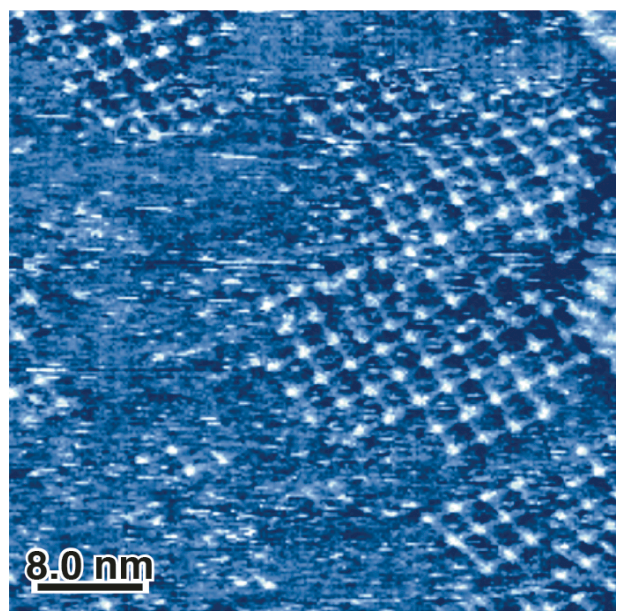

**Fig. 2.** Representative wide-area *in situ* STM image of covalently bonded FeTAPP–TPA mesh self-assembled on iodine-modified Au(111) in an aqueous solution.

High-resolution images of FeTAPP and $H_2$TAPP mesh frameworks, provided in Figure 3A and B, highlight the structural similarities and differences in image contrast between FeTAPP and $H_2$TAPP. Specifically, each FeTAPP molecule connected in the mesh exhibited greater contrast at its center due to an relative increase in tunneling current passing through the partially filled $dz^2$ orbital of the coordinated Fe ion.[54-56] The measured distance between neighboring FeTAPP molecules connected in an ideal mesh was 2.5±0.1 nm, showing strong agreement with our previous reports of the $H_2$TAPP.[47] Expected differences in inter-porphyrin distances were not successfully observed over experimental error for all combinations, $H_2$TAPP-TPA-$H_2$TAPP, FeTAPP-TPA-$H_2$TAPP and FeTAPP-TPA-FeTAPP, due to relatively large thermal drift at room temperature and conformational diversity. The structures of the FeTAPP–TPA and $H_2$TAPP meshes are therefore considered functionally equivalent. However, TPA linkers in the $H_2$TAPP–TPA mesh have conformational variations, either *E* or *Z*, indicating potential flexibility and diversity of the generated framework structure. The FeTAPP–TPA system (Figure 3C and C') tended to construct an ideal symmetrical square structure with a few irregular conformations. In contrast, the $H_2$TAPP–TPA system (Figure 3D and D') frequently produced a mesh structure including distorted trapezoids.

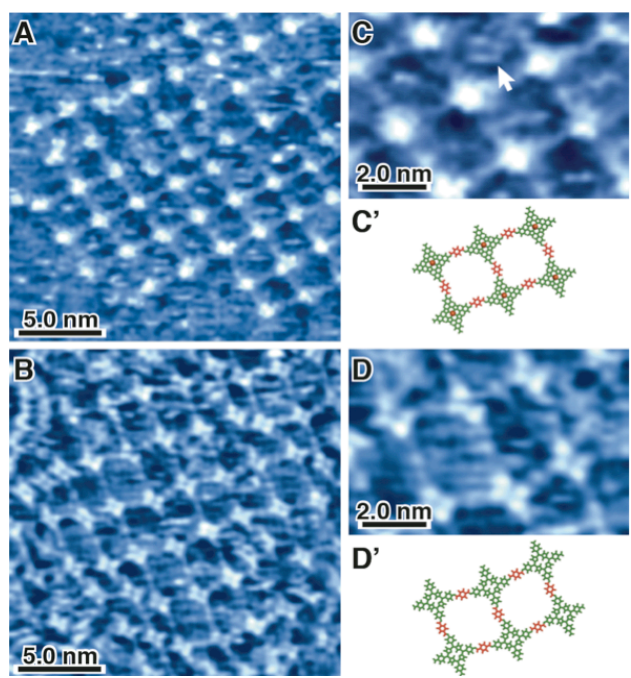

**Fig. 3.** Molecular-resolution *in situ* STM images (A–D), and corresponding models (C' and D') of FeTAPP–TPA (A, C) and $H_2$TAPP–TPA (B, D) meshes.

This behavior might be due to the effects of incorporated guest species, either TPA or FeTAPP, within the pores of the host mesh. The exact shapes of the various guest molecules were not highly resolved, likely as a result of rotational freedom and a high level of guest exchange. Small, irregularly shaped features were almost always observed in the mesh cavities, while TAPP shaped features trapped in the cavities were rarely observed as marked with an arrow in Fig. 3C. From visual inspection of the STM images, the appearance of guest molecules was seen to be higher in the FeTAPP–TPA system as compared to the $H_2$TAPP–TPA system at similar coverage, possibly due to host-guest interactions between the mesh, guest species, and additional counter-ions. This observation indicates a role for guest molecules in stabilization of the ideal symmetrical square mesh structure for the FeTAPP-TPA system, possibly due to electrostatic interactions between the mesh, guest species, and additional counter-ions.

*Corresponding author: kunitake@kumamoto-u.ac.jp

*3.2 Construction of ordered mesh structure form mixed FeTAPP–TPA and $H_2$TAPP–TPA systems*

Mixed FeTAPP–TPA and $H_2$TAPP–TPA hetero-molecular systems were also produced, using two methods: (1) simultaneous self-assembly and (2) partial replacement of the FeTAPP with $H_2$TAPP in an as-formed homo-molecular mesh. Yoshimoto and coworkers have reported highly ordered adlayers consisting of a prophyrin and a phthalocyanine arranged alternately.[57-59]

The mixed mesh structures were self-assembled from a solution consisting of FeTAPP, $H_2$TAPP, and TPA. These structures were observed across the entire surface area, as shown in Figure 4A. In this approach, a higher concentration of FeTAPP (approximately 3 to 10 times) than that of $H_2$TAPP was necessary to construct a mixed mesh structure. At equivalent solution concentrations, an $H_2$TAPP homo-mesh structure was observed exclusively and indicates that the content ratio (FeTAPP/$H_2$TAPP) in a mesh structure can be controlled by the solution content.

High-resolution images of the mixed mesh enabled identification and classification of porphyrin molecules located at each node as FeTAPP or $H_2$TAPP through their apparent height, as described above. FeTAPP and $H_2$TAPP were mixed at the molecular level, with hetero-connections between FeTAPP and $H_2$TAPP observed alongside homo-connections, as shown in Figure 4B and denoted by **M** (FeTAPP) and **H** ($H_2$TAPP). Intermolecular distances and shapes of the porphyrin structures indicate the formation of covalent bonds. Additionally, cross-sectional analysis along a lattice of a mixed mesh, shown in Figure 4C, revealed no difference in distances between porphyrins with homo-connections and those with hetero-connection.

A distributive location map of FeTAPP and $H_2$TAPP corresponding to Figure 4A is provided in Figure 4A', where red and blue circular spots, and dashed line between the spots indicate FeTAPP, $H_2$TAPP, and TPA connections, respectively. From this data, the molar percentages of FeTAPP and $H_2$TAPP were estimated to be 43% and 57%, respectively. In most of the mesh domains, the central region consisted almost exclusively of either FeTAPP (A) or $H_2$TAPP (B) with homo-connections (AA or BB), while hetero-connections (AB/BA) were mainly observed at the domain edges or boundaries. Although the molecularly mixed mesh lattice did not possess a regular connectivity pattern or alignment, the distribution of different molecule porphyrins was not completely random and showed a specific tendency. Through simple visual inspection, the ratio of homo-coupling (FeTAPP–FeTAPP and $H_2$TAPP–$H_2$TAPP) was obviously higher than that of hetero-coupling (FeTAPP–$H_2$TAPP). The average ratios of hetero- and homo-connections between $H_2$TAPP and FeTAPP molecules were 43% (AB/BA) and 57% (AA+BB = 34+23), respectively. These values are slightly shifted from the expected values of 49%, and 51% (33+19) calculated for random connection based on the observed molar percentages of each species respectively. The number of hetero- and homo-connections on each porphyrin shown in Figure 4D was clearly distorted from a random distribution toward fewer hetero-connections, thereby revealing a distinct preferential tendency toward the formation of homo-connections between adjacent porphyrins. These findings indicate that homo-coupling might be thermodynamically advantageous as compared with hetero-coupling, despite a minimal energy difference between the two structures.

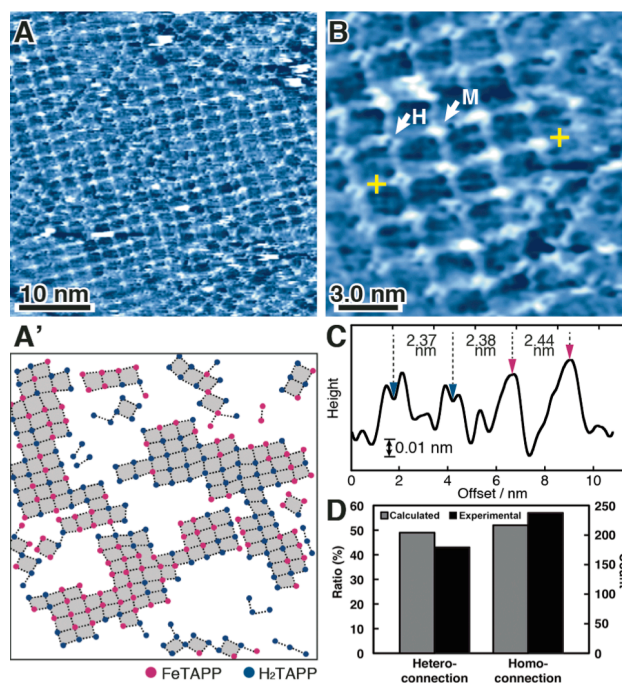

**Fig. 4.** Wide area (A) and zoomed (B) *in situ* STM images, corresponding distribution model (A') of FeTAPP and $H_2$TAPP, molecular cross-sectional profile (C) along a lattice in the mesh and histogram (D) of connecting unit distribution on each porphyrin. The crosses in (B) indicate start and end positions for the cross-section. In (D), gray and black bars show calculated and experimental abundance ratio (%) of homo- and hetero-connections on each porphyrin molecule, respectively.

To further examine observed prior indications of reaction selectivity between FeTAPP and $H_2$TAPP *via* TPA, partial replacement of the FeTAPP by $H_2$TAPP was performed following the construction and confirmation of an FeTAPP–TPA mesh that extended across the entire surface from a solution of FeTAPP and TPA (Figure 5A). Subsequent addition of $H_2TAPP$ resulted in a mesh structure in which FeTAPP was partially replaced by $H_2TAPP$. The final concentrations of FeTAPP, $H_2TAPP$, and TPA in the solution were 4.8, 0.51, and 30 µM, respectively. Figure 5B shows a representative STM image of mixed mesh framework acquired 20 min after the addition of $H_2TAPP$.

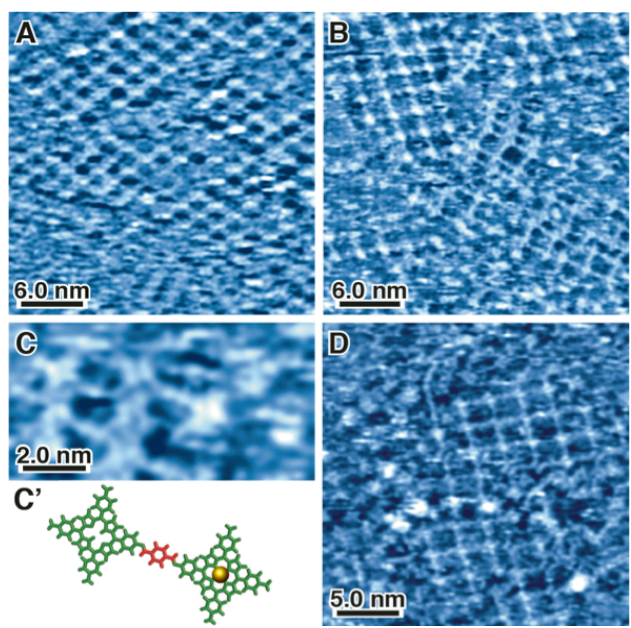

**Fig. 5.** Sequential in situ STM images of FeTAPP–TPA mesh before (A) and after (B and C) adding $H_2TAPP$ solution, the corresponding model (C'), and typical STM image observed several hours after addition of $H_2TAPP$ to FeTAPP-TPA mesh (D).

Coexistence of both FeTAPP–TPA and $H_2TAPP$–TPA frameworks was observed, as shown in the upper and lower right-hand regions of the image, respectively. Various irregular structures, including linear arrays, were commonly observed in $H_2TAPP$–TPA domains, while domain size was observed to decrease substantially following the addition of $H_2TAPP$. In some regions, smaller homo-domains of FeTAPP–TPA or $H_2TAPP$–TPA were also observed during the early stages of adsorption. At this stage, adsorption of $H_2TAPP$ in the inter-domain regions led to the construction of small $H_2TAPP$–TPA homo-mesh regions. Here, spontaneous adsorption and polymerization occurred rather than replacement of porphyrin molecules in the FeTAPP–TPA mesh. The final structure in which homo- and hetero-connections coexisted but where homo-connections dominated, observed several hours after $H_2TAPP$ addition, was essentially the same as that of the mesh structure prepared by simultaneous adsorption, as a result of continual molecular replacement (Figure 5D).

### 4. CONCLUSIONS and OUTLOOK

The capacity to generate 2-D π-conjugated metal–porphyrin covalent organic frameworks constructed by a soft solution methodology paves the way to true bottom-up assembly of a vast array of solid-supported, designer supramolecular nanoarchitectures toward applications including electronics, solar and fuel cells, biosensors, separations, nanoporous membranes, and commercial coatings as well as the potential to serve as surface templates for subsequent growth of extended 3-D architectures. Of particular note is the fact that azomethine coupling, applied here as "on-site" polycondensation, provides p-conjugated connections between building blocks, leading to expanded p-conjugated molecular systems. Apart from Ullmann coupling on a Cu substrate with heat treatment under UHV conditions, azomethine coupling is the only reaction among those reported for 2-D systems that gives p-conjugated connections. The construction of such frameworks, consisting of a mixture of functional building blocks, and the ability to identify their electronic properties at the molecular level represents an important step forward in the production of molecular devices and molecular circuits. A 2-D supramolecular arrangement comprised of a combination of porphyrins with different energy states could provide a simple biomimetic model for solar-energy-harvesting systems in terms of self-assembly. The preferential connections in mesh structures found here will be of significant use in designing more sophisticated covalent 2-D nanosystems with electronic communication. Visualization of local densities of states of p-conjugated polymeric architectures in real space will be exciting as a future work.

*Corresponding author: kunitake@kumamoto-u.ac.jp

### ACKNOWLEDGMENT

This work was partially supported by a Grant-in-Aid for Scientific Research on Innovative Areas "New Polymeric Materials Based on Element-Blocks" (24102006) of the Ministry of Education, Culture, Sports, Science and Technology, Japan.

*\*Corresponding author: kunitake@kumamoto-u.ac.jp*